\documentclass{PoS}

\renewcommand{\d}{\textmd{d}}
\newcommand{\Tr}{\textmd{Tr}}

\title{The curvature of the QCD phase transition line }

\ShortTitle{The curvature of the QCD phase transition line }

\author{\speaker{G.~Endr\H{o}di}$^1$, Z. Fodor$^{1,2}$, S.D.~Katz$^{1,2}$, K.K. Szab\'o$^2$\\
$^1$Institute for Theoretical Physics, E\"otv\"os University, H-1117 Budapest, Hungary. \\
$^2$Department of Physics, University of Wuppertal, D-42097 Wuppertal, Germany.\\
       E-mail: \email{endrodi@general.elte.hu}
}

\abstract{We determine the curvature of the phase transition line in the $\mu-T$ plane through an analysis of various observables, including the Polyakov loop, the quark number susceptibilities and the susceptibility of the chiral condensate. The second derivative of these quantities with respect to $\mu$ was calculated. The measurements were carried out on $N_T =4,6,8$ and $10$ lattices generated with a Symanzik improved gauge and stout-link improved 2+1 flavour staggered fermion action using physical quark masses.
}

\FullConference{The XXVI International Symposium on Lattice Field Theory \\
                July 14 - 19, 2008\\
                Williamsburg, Virginia, USA}

\begin{document}

\section{Introduction}

Quantum chromodynamics (QCD) is the theory of strong interactions. While at low temperatures the dominant degrees of freedom are hadrons, due to one of the most important properties of QCD, asymptotic freedom, at high temperatures it describes a different phase of matter called quark-gluon plasma (QGP). The phase transition between the hadronic phase of matter and QGP can be investigated by lattice simulations. The transition at zero chemical potential -- which represents the case of equal number of quarks and antiquarks -- is of huge importance, since it is relevant for the early Universe and also for high energy collisions. The region of the phase diagram corresponding to small and moderate $\mu$ is also interesting, since the cooling down of the QGP after a heavy ion collision occurs in this area.

Simulations at nonvanishing chemical potential are burdened by a significant problem: the fermion determinant here becomes complex, and as a result makes importance sampling impossible. A possible solution for this issue is to expand observables into a Taylor-series in $\mu$, where the coefficients can be calculated at vanishing chemical potential. Due to the symmetry of the partition function describing the system, the first term in these expansions always vanishes. The second term on the other hand is related to the curvature of the transition line.

The 2+1 flavour QCD transition was found to be an analytic crossover~\cite{Aoki:2006} (instead of a first-order phase transition), which usually results in different transition temperatures for different observables~\cite{Aoki:2006br}. Our aim is to determine the curvature $\kappa$ of the phase transition line for different observables. These include the Polyakov loop, the strange quark number susceptibility, the chiral condensate and the chiral susceptibility.

\section{Possible scenarios}

\begin{figure}[h!]
\centering
\mbox{
	\includegraphics*[height=6.4cm]{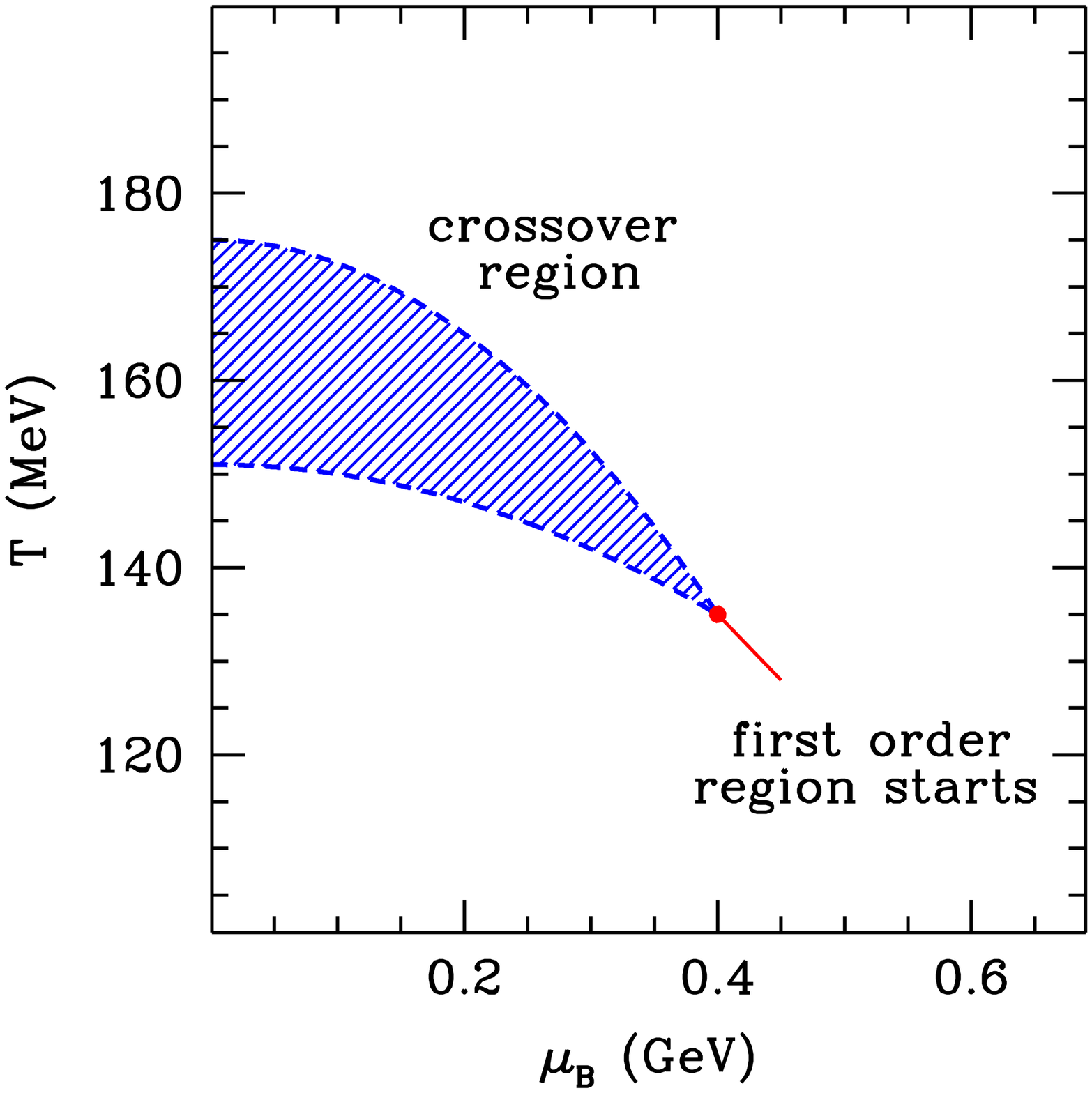}
     	\includegraphics*[height=6.4cm]{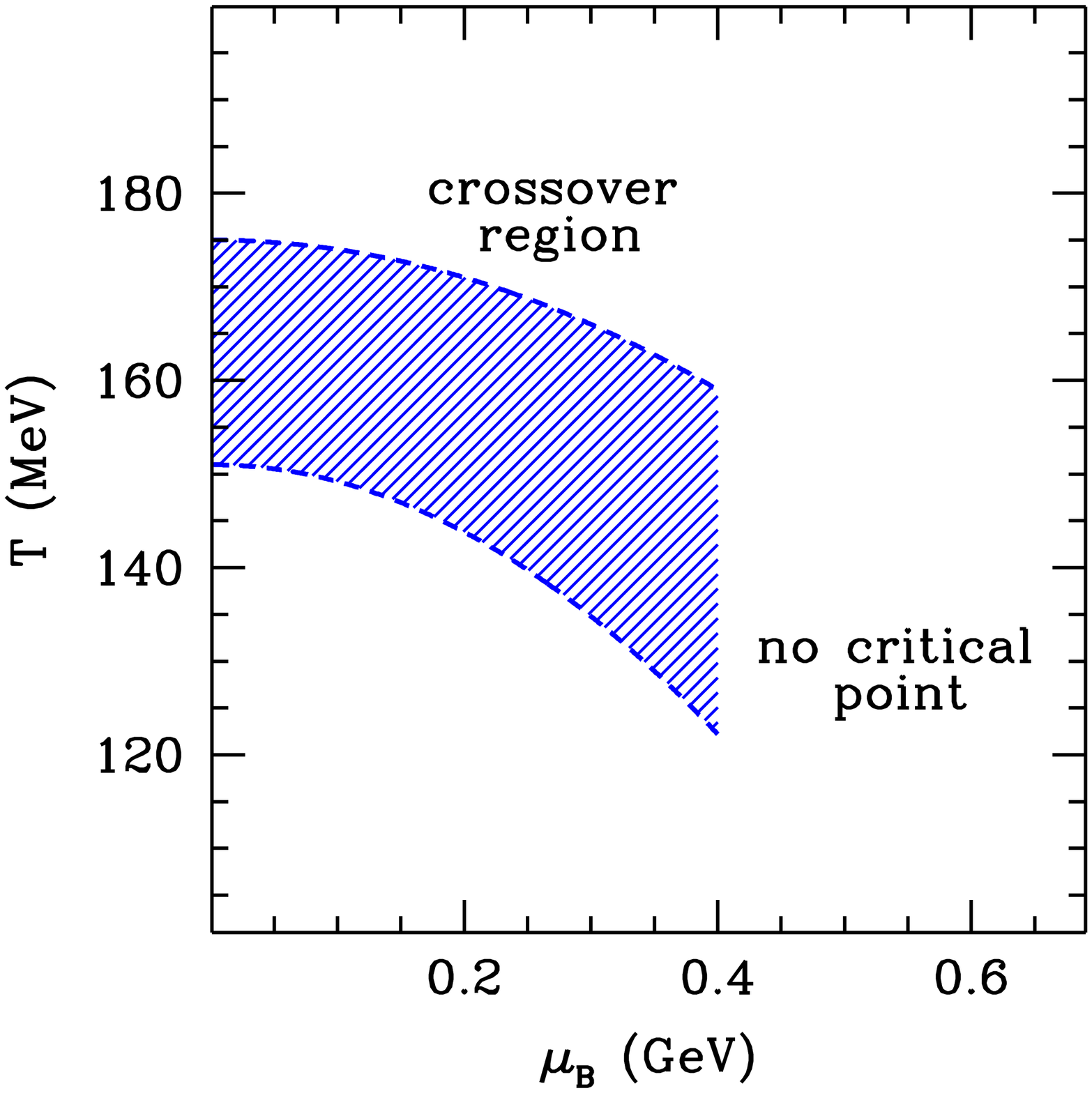}
}
\caption{Two possible scenarios about the QCD phase diagram around zero $\mu$. Transition temperatures determined by different observables may converge to (left side, compare to first lattice indications on $N_T=4$~\cite{Fodor:2001pe}) or diverge from (right side, first lattice indications on $N_T=4$~\cite{deForcrand:2007rq}) each other as the chemical potential is increased. This tendency can also have an effect on the existence (and if it exists, the position) of the QCD critical endpoint.}
\label{fig:scenarios}
\end{figure}

As an effect of the crossover character of the transition at zero $\mu$, the critical temperature determined from the chiral susceptibility ($T_c(\chi_{\bar\psi\psi}) \approx 151$ MeV) and that from the strange quark number susceptibility ($T_c(\chi_s) \approx 175$ MeV) are quite different~\cite{Aoki:2006br}. This situation can change at finite chemical potential according to two possible scenarios. The strengthening of the transition could cause these two values to approach each other, while if the transition remains a weak crossover, the two transition lines can diverge (see figure $\ref{fig:scenarios}$). The two cases are separated by the sign of the difference $\kappa(\chi_s)-\kappa(\chi_{\bar\psi\psi})$ of the curvatures for the two observables. A positive sign (left side) would indicate that the crossover region shrinks, which might suggest that a critical endpoint exists on the phase diagram. On the other hand a negative sign would cause the crossover region to expand, which may result in the abscence of the critical point.

Supposing that there exists a critical endpoint around baryonic chemical potential $\mu_B \approx 360$ MeV, the difference of the curvatures should be around $\Delta\kappa \approx 0.02$ (see definition below). Here and also in the following under $\mu$ we mean the $\mu_B$ baryonic chemical potential. There are several works in the literature in which the value of the curvature was computed~\cite{Allton:2002zi, de Forcrand:2002ci, D'Elia:2002gd, de Forcrand:2003hx, Fodor:2004nz}. In these $N_T=4$ works the curvature was found to be in the interval $\kappa=0.003 \ldots 0.01$. On the other hand, the $N_T=6$ results of~\cite{deForcrand:2007rq} indicate that the curvature might become less pronounced towards the continuum limit.
\section{The curvature}

Let us parametrize the transition line in the vicinity of the vertical $\mu = 0$ axis as $T_c(\mu)=T_c \left( 1 - \kappa \cdot \mu^2/T_c^2 \right)$. This implies that the curvature can be written as 

\begin{equation}
\kappa= - T_c \left. \frac{\d T_c(\mu)}{\d (\mu^2)} \right|_{\mu=0}
\end{equation}

We illustrate our first procedure to measure the derivative in question using the quantity $\chi_s/T^2$ (see definition in section~\ref{sec:defs}). Since $0 \le \chi_s/T^2 \le 1$ stands independently of $\mu$, we can define $T_c(\mu)$ via the relation $\chi_s/T^2 (T_c, \mu) = 0.5$. In the following we use the short notation $\mathcal{O} \equiv \chi_s/T^2$.

The total derivative of the observable $\mathcal{O}(T,\mu^2)$ may be written as $\d\mathcal{O}=\left (\partial \mathcal{O}/\partial T \right ) \cdot \d T + \left (\partial \mathcal{O}/\partial (\mu^2) \right )\cdot \d \mu^2$. So, since along the $T_c(\mu)$ line $\d\mathcal{O}=0$ by definition, for the derivative in the definition of the curvature this reads as

\begin{equation}
\frac{\d T_c}{\d \mu^2}= - \left( \frac{\partial \mathcal{O}}{\partial \mu^2} \right) \Big / \left(\frac{\partial \mathcal{O}}{\partial T} \right)
\label{eq:tcdef}
\end{equation}

This way it is enough to measure these two derivatives at an arbitrary temperature. For more complicated quantities this procedure gives only an approximate value for the curvature. In these cases one should carry out measurements at various temperatures; then $T_c(\mu)$ can be defined as the inflection point (for e.g. $\chi_s/T^2$) or the maximum value (for e.g. $\chi_{\bar\psi\psi}$) of the observable. This second procedure -- which is clearly more CPU-demanding -- is an ongoing project. Nevertheless, to calculate the derivative $\partial \mathcal{O}/\partial (\mu^2)$ we need to measure new operators. These are detailed in the next section.

\section{Technique}

Let us consider the partition function in its usual form

\begin{equation}
\mathcal{Z}=\int{\mathcal{D}U e^{-S_g(U)} (\det M)^{N_f/4}}
\end{equation}
\noindent
and denote the derivative with respect to $\mu_{u,d}$ by $'$. The derivatives of $\mathcal {Z}$ are easily calculated to be $(\log \mathcal{Z})'=\langle n_{u,d} \rangle$ and $(\log \mathcal{Z})''=\langle \chi_{u,d} \rangle$, where the light quark number density $n_{u,d}$ and the light quark number susceptibility $\chi_{u,d}$ are the following combinations:
\begin{eqnarray*}
n_{u,d} &=& \frac{N_f}{4}\Tr\left( M^{-1}M'\right)\\
\chi_{u,d} &=& n_{u,d}^2 + \frac{N_f}{4} \Tr\left( M^{-1}M' - M^{-1}M'M^{-1}M'\right)
\end{eqnarray*}
\noindent
Using these definitions the second derivative of any (possibly $\mu_{u,d}$-dependent) observable can be determined as

\begin{equation}
\frac{\partial^2 \langle \mathcal{O} \rangle}{\partial \mu_{u,d}^2}=\langle \mathcal{O} \chi_{u,d}\rangle - \langle \mathcal{O} \rangle \langle \chi_{u,d}\rangle +\langle 2\mathcal{O}'n_{u,d} + \mathcal{O}''\rangle
\label{eq:obsder}
\end{equation}
\noindent
For observables that do not depend explicitly on $\mu_{u,d}$ (like the Polyakov loop $L$ or $\chi_s$) the third term in (\ref{eq:obsder}) vanishes. For $\mu_{u,d}$-dependent observables the derivatives $\mathcal{O}'$ and $\mathcal{O}''$ were calculated numerically, using a purely imaginary chemical potential.

\section{Simulation setup}

We used a Symanzik improved gauge and stout-link improved staggered fermionic lattice action in order to reduce taste violation. The configurations were generated with an exact RHMC algorithm. We determined the line of constant physics (LCP) using physical masses for the light quarks $m_{u,d}$ as well as for the strange quark $m_s$. The LCP was fixed by setting the ratio $m_K/f_K$ and $m_K/m_\pi$ to their physical values. We used four different lattice spacings $N_T = 4, 6, 8, 10$ and aspect ratios $N_S/N_T$ of $4$ and $3$. The scale was fixed by $f_K$ and its unambiguity checked by calculating $m_{K∗}$, $f_\pi$ and $r_0$. For measuring the operators necessary for the above mentioned derivative the random noise estimator method was used. The number of random vectors was set to $80$, in order for the error coming from the random estimator method and that from the finiteness of the statistics to be of the same extent. The details of the simulation setup can be found in~\cite{Aoki:2006br} or~\cite{Aoki:2005vt}.

\section{Observables}
\label{sec:defs}

In order to determine the derivative $\partial T_c/\partial (\mu^2)$ expressed in (\ref{eq:tcdef}), one needs to select some observable $\mathcal{O}$. For this role we used the following quantities:

The Polyakov loop is defined as

\begin{equation}
L=\frac{1}{N_S^3}\sum_x \Tr \prod_{t=0}^{N_T-1}U_4(x,t)
\end{equation}
\noindent
In order to be able to extract continuum limit results, an appropriate renormalization is necessary. For $L$ this means $L_r=L \exp(V(r_0)/2T)$, where $V(r)$ is the $T=0$ static potential~\cite{Aoki:2006br}.

The strange susceptibility 
\begin{equation}
\chi_s=\frac{T}{V}\frac{\partial^2 \log \mathcal{Z}}{\partial \mu_{u,d}^2}
\end{equation}
\noindent
needs no renormalization, since it is connected to a conserved current.

The chiral condensate can also be expressed as a derivative of the partition function:
\begin{equation}
\bar\psi\psi=\frac{T}{V}\frac{\partial \log \mathcal{Z}}{\partial m}
\end{equation}
It can be renormalized by subtracting the additive divergences, and then multiplying by the quark mass, so that the multiplicative factors also cancel (here the factor $m_\pi^4$ is used to get a dimensionless combination):
\begin{equation}
\bar\psi\psi_r=(\bar\psi\psi - \bar\psi\psi(T=0))\cdot m \cdot \frac{1}{m_{\pi}^4}
\end{equation}
\noindent

Finally, the definition of the chiral susceptibility is
\begin{equation}
\chi_{\bar\psi\psi}=\frac{T}{V}\frac{\partial^2 \log \mathcal{Z}}{\partial m^2}
\end{equation}
\noindent
and the renormalization is the following:
\begin{equation}
\chi_{\bar\psi\psi r}=(\chi_{\bar\psi\psi} - \chi_{\bar\psi\psi}(T=0))\cdot m^2 \cdot \frac{1}{T^4}
\end{equation}

\section{Results}

\begin{figure}[h!]
\centering
\mbox{
	\includegraphics*[height=6.2cm]{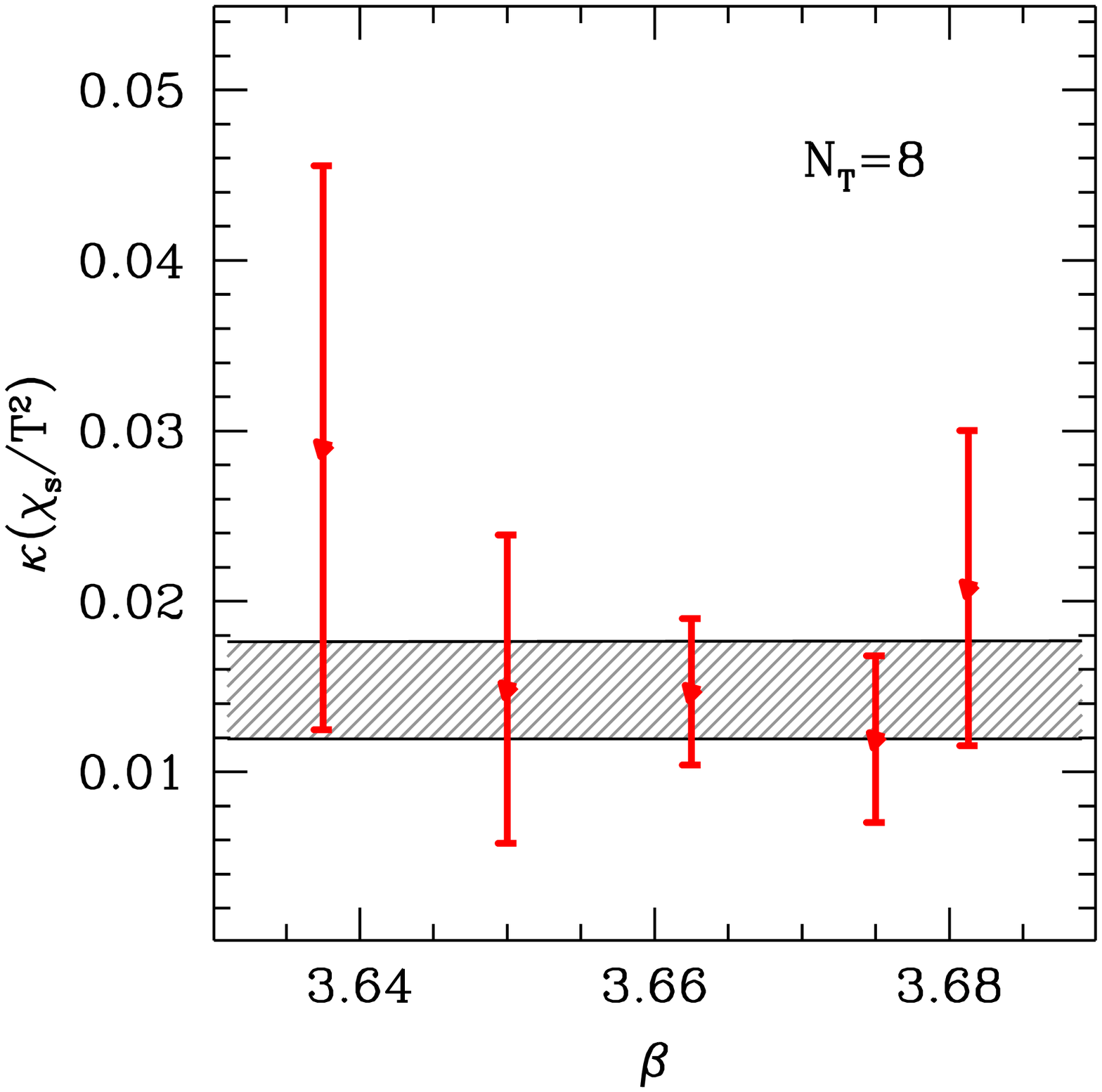}
     	\includegraphics*[height=6.2cm]{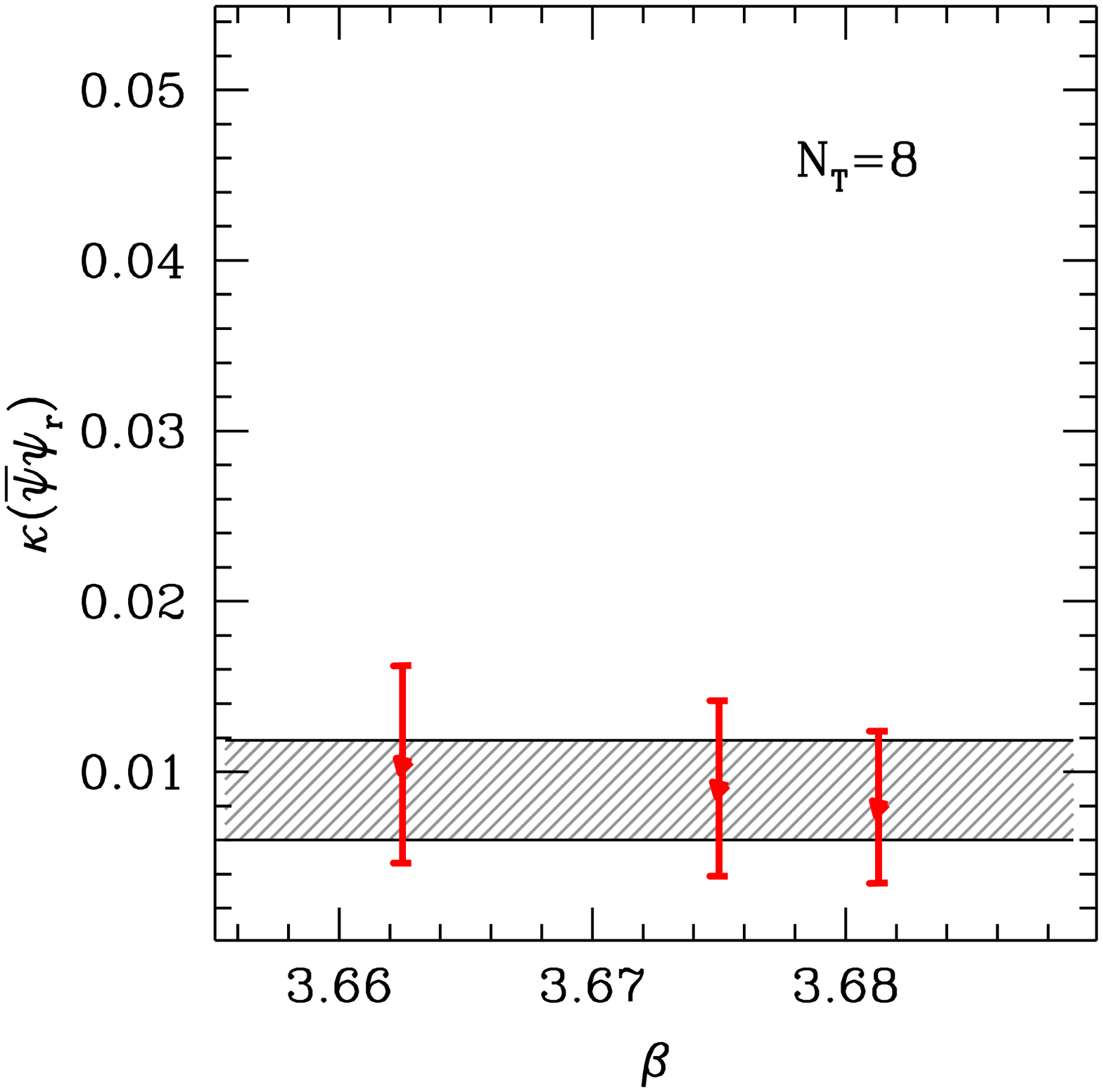}
}
\caption{The curvature determined using the normalized strange susceptibility (left side) and the renormalized chiral condensate (right side), respectively. Independent measurements at different temperatures ($\beta$ values) can be averaged to decrease statistical errors. Results from $N_T=8$ lattices are shown.}
\label{fig:avg}
\end{figure}

As already mentioned, in our approach we select an arbitrary temperature to calculate the curvature of the transition line. In order to increase statistics, we can also average measurements at different temperatures. This procedure is illustrated on figure~\ref{fig:avg} for the case of the strange susceptibility and the chiral condensate.

\begin{figure}[h!]
\centering
\mbox{
	\includegraphics*[height=7.0cm]{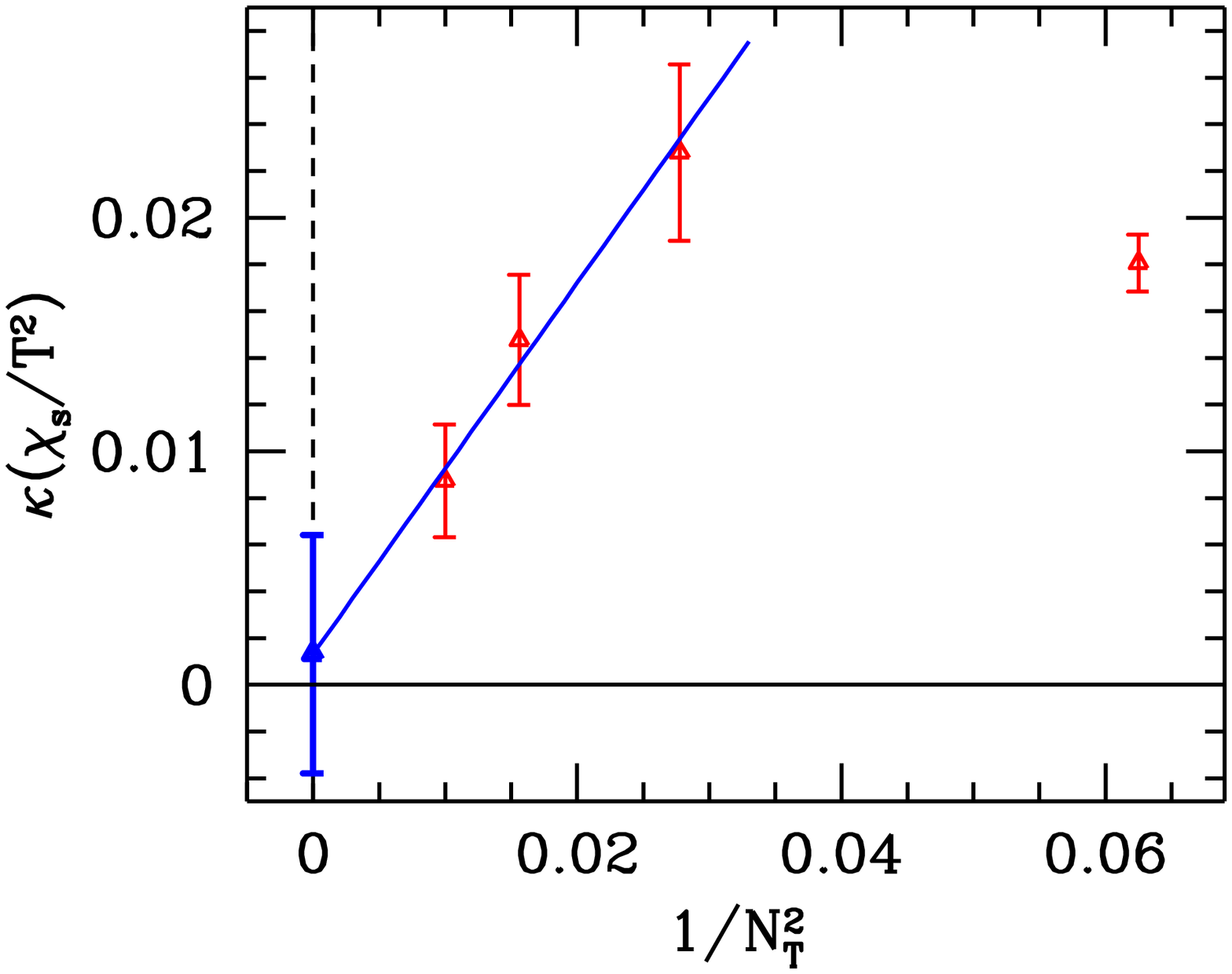}
     	\includegraphics*[height=7.0cm]{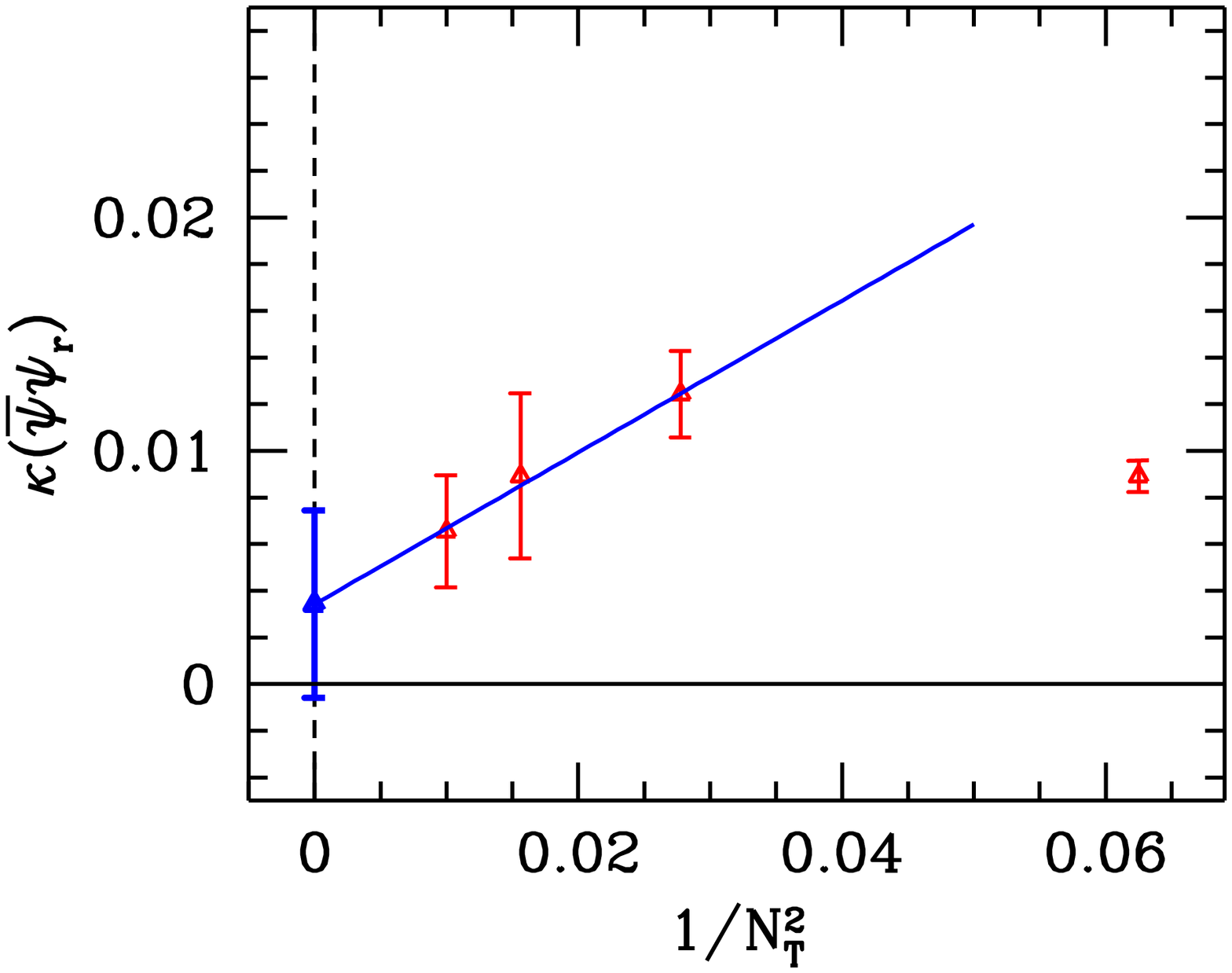}
}
\caption{The curvature of the transition line as a function of the lattice spacing squared, for the strange susceptibility (left side) and the chiral condensate (right side). As characteristic to the applied action, the $N_T=4$ results are out of the scaling regime, so the linear extrapolation is carried out using only the finer lattices. As it can be seen, larger statistics is needed in order to extract a reliable continuum result.}
\label{fig:cont}
\end{figure}

Having obtained the renormalized quantities for $N_T=4,6,8$ and $10$, we are in a position to carry out the continuum extrapolation. The results shown on figure~\ref{fig:cont} are only preliminary; note that the extrapolated values have quite large statistical errors. 

\section{Summary}

While the curvature values for $\bar\psi\psi_r$ are smaller than those for $\chi_s/T^2$ at every lattice spacing, they become consistent in the $a \rightarrow 0$ limit. Our aim in the first place is to determine the difference $\kappa(\bar\psi\psi_r)-\kappa(\chi_s/T^2)$, so increasing the statisctics in order to clarify the situation is very important. Nevertheless we report the continuum results: $\kappa(\bar\psi\psi_r)=0.0034(40)$ and $\kappa(\chi_s/T^2)=0.0013(51)$, and also for the Polyakov loop and the chiral susceptibility: $\kappa(L_r)=-0.0095(93)$ and $\kappa(\chi_{\bar\psi\psi r})=-0.0018(34)$.

As the next step we are looking forward to perform the above described procedure to measure $\partial \mathcal{O}/\partial (\mu^2)$ for the whole temperature interval around $T_c$. This way it becomes possible to determine the critical temperature for the given observables (by locating their inflection point or maximum point) at nonvanishing chemical potential. Using this technique and much larger statistics we plan to give a final answer for the curvature of the QCD transition line.

\end{document}